\documentclass[%
 reprint,
amsmath,amssymb,
 aps,
superscriptaddress]{revtex4-1}

\usepackage{book tabs}
\usepackage{graphicx}
\usepackage{dcolumn}
\usepackage{bm}
\usepackage{relsize}
\usepackage{float}
\usepackage{adjustbox}
\usepackage{color}
\usepackage[normalem]{ulem}
\usepackage{appendix}
\usepackage{tensor}

\usepackage{color}
\definecolor{darkgreen}{rgb}{0.34, 0.55, 0.23}

\begin{document}

\preprint{APS/123-QED}

\title{Positivity bounds on reconstructed Horndeski models}

\author{Joe~Kennedy}
\author{Lucas~Lombriser}
\affiliation{D\'{e}partement de Physique Th\'{e}orique,
Universit\'{e} de Gen\`{e}ve,
24 quai Ernest Ansermet,
1211 Gen\`{e}ve 4,
Switzerland}

\date{\today}

\begin{abstract}
Positivity bounds provide conditions that a consistent UV-completion exists for a quantum field theory. We examine their application to Horndeski gravity models reconstructed from the effective field theory (EFT) of dark energy.
This enables us to assess whether particular phenomenological parameterizations of the EFT functions reconstruct theories that respect or violate the positivity bounds.
We find that commonly adopted EFT parametrizations, cast in terms of the dark energy density or power laws of the scale factor,
only satisfy the positivity bounds in non-trivial regions of the parameter space.
We then examine parameterizations of the inherently stable EFT basis, constructed to avoid gradient and ghost instabilities by default.
In stark contrast, in this basis the positivity bounds either only provide constraints in \emph{a-priori} unrealistic regions of the parameter space or do not provide any constraints on parameter values at all.
The application of positivity bounds to common parametrizations of the standard EFT functions can therefore lead to artificial conclusions that the region of viable Horndeski modifications of gravity is highly constrained.
Our results provide a strong motivation, in addition to the default avoidance of theoretical instabilities, for instead adopting parametrizations of the inherently stable EFT basis when testing dark energy and modified gravity models with forthcoming cosmological survey data.
\end{abstract}

\pacs{Valid PACS appear here}
\maketitle

\section{Introduction} \label{sec:intro}


Understanding the physical nature of the late-time accelerated expansion of the Universe~\cite{Riess:1998cb, Perlmutter:1998np} is a central endeavour of cosmological research, which has the potential to make a significant impact on both our understanding of cosmology and fundamental physics.
It is therefore an ongoing effort to develop theoretical models which can drive accelerated expansion as well as leave traces in the large-scale structure.
Even disregarding cosmic acceleration, these models provide a playground with which to test extensions to General Relativity (GR) in the cosmological regime.
Adding a classical cosmological constant to GR provides the most straightforward explanation of cosmic acceleration, but a number of theoretical difficulties arising from quantum corrections, well known even prior to the discovery of the accelerated expansion~\cite{Weinberg1989}, led theorists to consider alternatives~\cite{Martin:2012bt, Burgess:2013ara, Padilla:2015aaa}.
Perhaps the simplest extension beyond a cosmological constant is to incorporate an additional scalar field into the Einstein-Hilbert action which acts as the driver of accelerated expansion.
It can enter as an additional component in the stress-energy content of the Universe, as a modification of GR through a non-minimal coupling function or a combination of the two \cite{Joyce:2014kja, Joyce:2016vqv}.
This scalar field may be considered to be a new fundamental field or arising as a low energy effective remnant of an as yet unknown UV-complete theory \cite{Fujii:2003pa}.

Requiring that the theory results in at most second-order equations of motion ensures that the resulting Hamiltonian is bounded from below in accordance with Ostrogradsky's theorem~\cite{Woodard:2006nt}. 
Ensuring second-order equations of motion restricts the resulting scalar-tensor theory to be the Horndeski action~\cite{Horndeski:1974wa, Deffayet:2011gz, Kobayashi:2011nu}.
It is also possible to construct stable theories that have higher-order equations of motion and avoid the non-degeneracy condition of Ostrogradsky's theorem~\cite{Gleyzes:2014dya, Langlois:2015cwa}.
We restrict solely to Horndeski theories in this work.

The simultaneous detection of gravitational waves with an electromagnetic counterpart constrained the deviation of the speed of gravitational waves from the speed of light to $\mathcal{O}(10^{-15})$~\cite{TheLIGOScientific:2017qsa}.
Tight constraints could then be placed on the remaining freedom in Horndeski theory~\cite{Lombriser:2015sxa, Lombriser:2016yzn, McManus:2016kxu, Baker:2017hug, Sakstein:2017xjx, Ezquiaga:2017ekz, Creminelli:2017sry},
modulo some caveats~\cite{Battye:2018ssx, deRham:2018red, Copeland:2018yuh}. 
%
%
%
Even in light of this tight constraint, significant freedom remains in Horndeski theory. 
A powerful tool that is commonly employed to study the effects Horndeski theory has on cosmological perturbations is the Effective Field Theory (EFT) of dark energy~\cite{Creminelli:2008wc, Park:2010cw, Bloomfield:2011np, Gubitosi2012, Bloomfield:2012ff, Gleyzes:2013ooa, Tsujikawa:2014mba, Gleyzes:2014rba, Bellini:2014fua, Lagos:2016wyv, Frusciante:2019xia}.
This enables one to encapsulate the effects of Horndeski theory on the background expansion and the cosmological perturbations in terms of a set of free functions of time, commonly called the EFT parameters. 
Different basis descriptions for the EFT parameters each have their own useful properties \cite{Gubitosi2012, Bellini:2014fua}.
A particularly useful basis which naturally avoids ghost and gradient instabilities, associated with kinetic terms with the wrong sign and divergent scalar field clustering, was developed in Refs.~\cite{Kennedy:2018gtx, Lombriser:2018olq}. 
By restricting to this basis of EFT functions, it is guaranteed that any sampling of the parameter space automatically satisfies these stability conditions.

With the aim of testing Horndeski theory with cosmological data-sets, there are two possible approaches. 
One can pick a favored Horndeski theory and compute the resulting EFT parameters.
However, this can be considered ad-hoc as one is biased towards the particular chosen theory. 
By contrast, starting with EFT allows for a generic description of the imprints of Horndeski theory on the cosmological background and perturbations without the need to restrict to a fixed model. 
Studying the effects on cosmological observables, such as the linear matter power spectrum, for a given set of EFT parameters is made possible with modern codes which compute cosmological anisotropies in dark energy and modified gravity models such as \emph{hi\_class}~\cite{Zumalacarregui:2016pph}.
The drawback with this generic approach is that it is not obvious which underlying Horndeski theory is being tested with a particular choice of EFT parameters. 
Motivated by this issue Refs.~\cite{Kennedy:2017sof, Kennedy:2018gtx} developed a reconstruction which maps a given set of EFT parameters to the class of underlying Horndeski models they correspond to.  
With this mapping it is possible to connect results that apply at the level of the Horndeski action with results on EFT parameters and vice-versa.
For example, using this approach Ref.~\cite{Kennedy:2019nie} demonstrated how to implement screening mechanisms in theories that are only constrained at the level of the linear perturbations.


Connecting theoretical methods that apply at the level of the covariant action to those at the level of EFT proves particularly useful when studying how to place priors arising from theoretical arguments on the parameter space.
By restricting the parameter space to regions that satisfy desired theoretical priors it is not necessary to verify each sampled point satisfies the theoretical condition, increasing the computational efficiency of maximum likelihood analyses. 
There has already been much work in exploring the utility novel theoretical arguments provide in constraining the viable dark energy and modified gravity model space, such as Refs.~\cite{Creminelli:2018xsv, Creminelli:2019kjy}, which study the impact of instabilities induced by scalar-graviton interactions.
A further advantage of studying theoretical priors is that they provide a way to use observational constraints to directly test the underlying assumptions on which they based.
%


A particularly interesting prior that one could place on the EFT parameter space comes from the study of \emph{positivity bounds} in quantum field theory. 
These are conditions that can be placed on the signs of certain parameters in the low-energy effective theory in order for the high-energy theory to respect standard properties of quantum field theory such as locality, causality and crossing symmetry.
Their power lies in the fact that one can remain ignorant about the precise form of the high-energy theory, while still placing bounds on the coefficients of operators in the effective theory.
With regards to cosmology, Ref.~\cite{Melville:2019wyy} demonstrated how it is possible, if Horndeski theory is to be regarded as a low-energy remnant from some unknown UV-complete theory, to utilise positivity bounds to place constraints on the free functions in the theory.
%

%
In this paper we tackle the problem from an alternative perspective by combining the reconstruction method of Refs.~\cite{Kennedy:2017sof, Kennedy:2018gtx} with the positivity bounds derived for Horndeski theory in Ref.~\cite{Melville:2019wyy}. 
We study popular parameterizations of the EFT functions and determine whether the resulting reconstructed theory satisfies the positivity bounds across the parameter space. 
Put another way, we determine whether the underlying Horndeski theory of a given set of EFT functions parameterised in a particular basis satisfies the positivity bounds.
We then also study the region in the parameter space that satisfies the positivity bounds when the model is parameterised in terms of the inherently stable basis of Refs.~\cite{Kennedy:2018gtx, Lombriser:2018olq}.

%
The paper is organised as follows. In Sec.~\ref{sec:background} we discuss the necessary background of the reconstruction method and the concept of positivity bounds.
In Sec.~\ref{Sec:III} we combine the reconstructed action with the positivity bounds placed on Horndeski theory in order to impose positivity priors on the EFT parameter space.
In Sec.~\ref{Sec.IV} we then employ popular parameterizations of the EFT functions and examine how imposing the requirement of positivity impacts the viable parameter space.
We then examine in Sec.~\ref{Sec.V} the impact on a theory parameterized in terms of the inherently stable basis from applying the positivity bounds.
Our conclusions are presented in Sec.~\ref{Sec:conclusions}.

\section{Reconstructed Horndeski models and positivity bounds} \label{sec:background}

The most general scalar-tensor theory with at most second-order equations of motion and speed of gravitational waves equal to the speed of light is given by the following Horndeski model~\cite{McManus:2016kxu},
\begin{equation}
    \sum_{i=2}^{4} \, \int d^{4}x \sqrt{-g} \, \mathcal{L}_{i} \, ,
\end{equation}
where 
\begin{eqnarray}
 \mathcal{L}_{2} & \equiv & G_{2}(\phi,X) \,, \label{Horndeski2} \\
 \mathcal{L}_{3} & \equiv & G_{3}(\phi, X)\Box \phi \,, \label{Horndeski3} \\
 \mathcal{L}_{4} & \equiv & G_{4}(\phi) R\,. \label{Horndeski4}
\end{eqnarray}
Each $G_{i}(\phi,X)$ with $i=2,3$ are free functions of the scalar field and its kinetic term $X\equiv \partial_{\mu}\phi\partial^{\mu}\phi$ whereas $G_{4}(\phi)$ is restricted to solely a function of $\phi$ and describes a non-minimal coupling to GR by multiplying the Ricci scalar.
The dynamics of the cosmological perturbations of this theory can be obtained by utilising the framework of the EFT of dark energy~\cite{Bloomfield:2013efa, Gleyzes:2014rba, Bellini:2014fua}.
This is formulated in the unitary gauge, where the spacetime is foliated with uniform time hypersurfaces as in the ADM formulation of GR \cite{Arnowitt:1962hi} with the scalar field perturbations being absorbed into the metric. 
Having broken time diffeomorphism invariance but retained spatial diffeomorphism invariance, the EFT is constructed with the cosmological perturbations acting as the operators, and every combination of operators consistent with broken time diffeomorphisms appears in the action.

More precisely, the EFT is constructed using the time-time component of the metric $\delta g^{00}$ and the extrinsic curvature tensor $K_{\mu\nu}$ of the constant time hypersurfaces, and traces thereof, as the operators with additional free functions of time which couple to them. 
Note there are further operators one could include, but we neglect these as they are only relevant for describing theories beyond those in the class of Eqs.~\eqref{Horndeski2}--\eqref{Horndeski4}. 
The EFT action that describes their background and linear dynamics takes the following form 
\begin{align}
S = & \: S^{(0,1)}+S^{(2)}+S_{M}[g_{\mu\nu},\Psi_m] \,,
\label{eftlag}\\
S^{(0,1)} = & \: \frac{M_{*}^{2}}{2}\int d^{4}x \sqrt{-g} \left[ \Omega(t) R -2\Lambda(t)-\Gamma(t)\delta g^{00} \right] \,,
\label{eq:s01} \\
S^{(2)} = & \frac{1}{2} \int d^{4}x \sqrt{-g} \left[ M^{4}_2(t)(\delta g^{00})^2-\bar{M}^{3}_{1}(t) \delta K \delta g^{00}\right]  \,.
\label{eq:s2}
\end{align}
Note that the perturbation of the extrinsic curvature tensor does not appear in $S^{(0,1)}$ as it is equivalent to what already appears up to a total derivative.
The coefficients $\left\{ \Omega(t),\Gamma(t),\Lambda(t),M_{2}^{4}(t),\bar{M}_{1}^{3}(t) \right\}$~\cite{Lombriser:2014ira, Kennedy:2017sof} are dimensionful parameters, whose precise functional form depends upon the choice of the $G_{i}$ functions from which they are derived.

If the background expansion is assumed to be $\Lambda$CDM the two Friedmann equations reduce the freedom to three free functions that can affect the linear perturbations.
It is possible to translate between this EFT basis and others which provide some physical intuition between the properties of each of the three functions.  
For example, Ref.~\cite{Bellini:2014fua} demonstrated that it was possible to express the three free functions which modify the linear perturbations as $\left\{ \alpha_{M},\alpha_{B},\alpha_{K} \right\}$ describing a possible evolution of the squared Planck Mass $M^{2}$, braiding, and kineticity respectively.
This basis has been employed in constraining EFT with data~\cite{Bellini:2015xja, Ade:2015rim} and we shall make use of it in Sec.~\ref{Sec.IV}.
Alternatively, Refs.~\cite{Kennedy:2018gtx, Lombriser:2018olq} proposed a different basis parameterised by $\left\{ M^{2}, c_{s}^{2},\alpha \right\}$, the Planck Mass, the soundspeed of the scalar field perturbations and the kinetic energy of the scalar field perturbations.
This basis has the advantages that there are definite priors which can be placed on these parameters, namely $M^{2}>0$, $0<c_{s}^{2}<1$ and $\alpha>0$, which arise from the requirement that, along with the obvious fact the Planck Mass squared be positive, the theory is also free of gradient and ghost instabilities.
By working in this basis, one is only encompassing models that are free from these instabilities, removing the need to perform stability checks on each point in the parameter space, albeit at the expense of solving a differential equation arising from the mapping between $c_{s}^{2}$ and $\alpha_{B}$ (see Sec.~\ref{Sec.V}). 
Furthermore, this basis avoids inefficient sampling due to $\Lambda$CDM being confined to a narrow corner of the parameter space, which occurs when using alternate bases.
%

%
Testing a specific Horndeski model against observations is a highly inefficient approach considering the essentially infinite amount of freedom in the theory.
Therefore an alternative method is commonly employed in order to constrain Horndeski theory with data, which assigns the EFT functions a general but phenomenologically motivated parameterization such that their effects only become relevant at late times. 
The gains in generality unfortunately sacrifice information on the underlying theory. 
Motivated by this issue Refs.~\cite{Kennedy:2017sof, Kennedy:2018gtx} developed a reconstruction which maps from a given set of EFT parameters back to the set of Horndeski theories that gives rise to the original generic parameterization.
Concretely, given a set of background and linear EFT parameters the reconstructed Horndeski theory is given by
\begin{align}
G_{2}(\phi, X) = & -M_{*}^{2}U(\phi) - \frac{1}{2}M_{*}^{2} Z(\phi)X+a_{2}(\phi)X^{2} \nonumber\\
 &+\Delta G_{2} \,,
\label{eq:G2recon} \\
G_{3}(\phi,X) = & \: b_{0}(\phi)+b_{1}(\phi)X+\Delta G_{3} \,,
\label{eq:G3recon} \\
G_{4}(\phi, X) = & \: \frac{1}{2}M_{*}^{2}F(\phi) \,,
\label{eq:G4recon}
\end{align}
where $U(\phi)$, $Z(\phi)$, $a_{2}(\phi)$, $b_{1}(\phi)$ and $F(\phi)$ are smooth functions of the EFT parameters, whose explicit expression depends on the order to which one has performed the reconstruction.
At the background and linear level they are given by \cite{Kennedy:2017sof}
\begin{eqnarray}
    U(\phi) & = &\Lambda + \frac{\Gamma}{2} - \frac{M_{2}^{4}}{2M_{*}^{2}} - \frac{9H\bar{M}_{1}^{3}}{8M_{*}^{2}} - \frac{(\bar{M}_{1}^{3})^{\prime}}{8} \, ,\\
    Z(\phi) & = & \frac{\Gamma}{M_{*}^{4}} - \frac{2M_{2}^{4}}{M_{*}^{6}} - \frac{3H\bar{M}_{1}^{3}}{2M_{*}^{6}} + \frac{(\bar{M}_{1}^{3})^{\prime}}{2M_{*}^{4}} \, , \\
    a_{2}(\phi) & = &\frac{M_{2}^{4}}{2M_{*}^{8}}+\frac{(\bar{M}^{3}_{1})^{\prime}}{8M_{*}^{6}}-\frac{3H\bar{M}_{1}^{3}}{8M_{*}^{8}} \, ,\\
    b_{1}(\phi) & = & \frac{\bar{M}_{1}^{3}}{2M_{*}^{6}} \, , \\
    F(\phi) & = & \Omega \, ,
\end{eqnarray}
where the primes in this case denote derivatives with respect to $\phi$.
Note that, as we restrict to models with luminal speed of gravitational waves, the only difference between the reconstructed theory in this paper and that of Ref.~\cite{Kennedy:2017sof} is that here we set $\bar{M}_{2}^{2}=0$.
The $\Delta G_{i}$ terms are corrections that one can add to $G_{2}$ and $G_{3}$ in order to move to another set of $G_{i}$ functions whose linear and background dynamics are equivalent. Their precise form is given by
\begin{equation}
    \Delta G_{2,3} = \sum_{n>2} \xi^{{\scriptscriptstyle(2,3)}}_{n}(\phi)\left(1+\frac{X}{M_{*}^{4}}\right)^{n} \, ,
    \label{DeltaG23}
\end{equation}
where the set of functions $\left\{ \xi^{{\scriptscriptstyle(2,3)}}_{n}(\phi) \right\}$ are completely free in the absence of nonlinear information. They can be used to, for example, implement screening mechanisms in the reconstructed theory \cite{Kennedy:2019nie}. 
The key property they possess in the context of this paper is that they vanish on the background solution for the scalar field $\phi=t M_{*}^{2}$ as here $X=X_{0}=-M_{*}^{4}$.

%
A further theoretical prior that one might wish to implement is that arising from the study of positivity bounds in quantum field theory, a subject which has been the focus of much research \cite{Adams:2006sv, Nicolis:2009qm, Bellazzini:2014waa, Baumann:2015nta, deRham:2017avq, Bellazzini:2017fep, Bellazzini:2019xts}.
These bounds fix the sign of coefficients in certain scattering amplitudes in the IR limit of a theory with a, possibly unknown, UV-completion. 
Importantly, it is not necessary to know the precise form of the UV-complete theory in order to utilise these positivity bounds in constraining the IR theory. 
The application of positivity bounds to the full Horndeski theory was done in Ref.~\cite{Melville:2019wyy}, which obtained bounds on the $G_{i}(\phi,X)$ functions, albeit in the class of theories whose tensor modes may not be luminal, as well applying them to existing cosmological data-sets.
In Sec.~\ref{Sec:III} we introduce the complementary approach of this work. Combining the reconstruction method with the positivity bounds derived for Horndeski theory we derive a new bound on the EFT parameters, before examining in Sec.~\ref{Sec.IV} and Sec.~\ref{Sec.V} the impact imposing this bound has on the available EFT parameter space.

%


\section{Positivity bound on EFT parameters from a recontructed Horndeski model }\label{Sec:III}
We shall now discuss in detail the approach of this paper.
The positivity bounds derived for Horndeski theory in Ref.~\cite{Melville:2019wyy} are applicable for a given set of the Horndeski $G_{i}$ functions.  
With a preferred model one can check whether there is a consistent UV-completion that satisfies what we expect of a healthy quantum field theory.
Constraints from cosmological surveys will take a different approach by utilising a generic set of parameters that make minimal assumptions about the underlying theory which lead to deviations from GR.
It is therefore of interest to examine whether the positivity bounds from Ref.~\cite{Melville:2019wyy} can be translated into bounds on more generic parameterizations. 
Ref.~\cite{Melville:2019wyy} derived a set of positivity bounds on the Horndeski functions that also encompass models with non-luminal speed of gravitational waves.
If we consider the $\phi\phi \rightarrow \phi\phi$ scattering process the amplitude is given by
\begin{equation}
    \mathcal{A}(s,t)=c_{ss}\frac{s^{2}}{\Lambda_{2}^{4}}+c_{sst}\frac{s^{2}t}{\Lambda_{3}^{6}} \, ,
\end{equation}
where $s$ and $t$ are Mandelstam variables describing the centre of mass energy of the scattering process and the momentum transfer between the particles. 
The quantities $\Lambda_{2}$ and $\Lambda_{3}$ are two energy scales characterising where the theory becomes strongly coupled, occurring at whichever scale is lower. 
Under the assumption that $\Lambda_{2}>\Lambda_{3}$ the positivity bounds require that $c_{ss}>0$ and $c_{sst}>0$.
Restricted to the subclass of models with luminal speed of gravity, $c_{sst}$ reduces to the trivial inequality that $G_{3}^{2}>0$ and so for our purposes we shall only make use of the condition on $c_{ss}$. Noting the different convention in the definition of $X$, this can be expressed as 
\begin{align}
    c_{ss}^{\phi\phi}&=G_{2XX}-G_{2\phi\phi}G_{3X}^{2}
    -2G_{3X}(G_{3\phi\phi}-G_{2\phi X}) >0 \,.
    \label{eq:Pos1}
\end{align}
It is expected that this result should hold for any smooth background evolution of $\phi$, in particular on a cosmological background.
One perspective to see this is to assert that Eq.~\eqref{eq:Pos1} is a bound on an effective theory in a given background. Its structure is determined by the requirement that there exists a healthy UV-completion for the EFT, a condition which should not be background dependent at the level of the EFT. 
Therefore in this paper we set the scalar field evolution to be equal to that of the cosmological time $\phi=tM_{*}^{2}$, as in the ADM-foliation of spacetime. 
Given this choice, the background evolution for the kinetic term of the scalar field is given by $X_{0}=-\dot{\phi}^{2}=-M_{*}^{4}$.
The functional form of the reconstruction is derived such that, once this identification is made and it is expanded in the perturbations, the action in Eqs.~\eqref{eq:s01} and \eqref{eq:s2} is recovered. 
Therefore setting $X_{0}=-\dot{\phi}^{2}=-M_{*}^{4}$ is a solution to the background equations of motion of the reconstructed theory. 
Furthermore, this background evolution of $\phi$ ensures the nonlinear correction terms in Eq.~\eqref{DeltaG23} vanish. 

We are now able to plug the reconstructed Horndeski functions in Eqns.~\eqref{eq:G2recon} to $\eqref{eq:G4recon}$ into Eq.~\eqref{eq:Pos1} in order to obtain the positivity bound expressed in terms of the EFT parameters. The positivity bound becomes
\begin{widetext}
\begin{align}
&\frac{H}{8 M_{*}^{16}} \left[ 2M_{*}^{8} \left(\bar{M}_{1}^{3}\right)^{\prime}-3\left(\bar{M}_{1}^{3}\right)^{3}\left((H^{\prime})^{2}+H H^{\prime\prime} \right)+2M_{*}^{4}\bar{M}_{1}^{3}\left(2H^{\prime}(\bar{M}_{1}^{3})^{\prime} -3M_{*}^{4}-8(M_{2}^{4})^{\prime}+2M_{*}^{2}\Gamma^{\prime}+2H(\bar{M}_{1}^{3})^{\prime\prime} \right)   \right] \nonumber \\ 
& +\frac{H (\bar{M}_{1}^{3})^{2}}{8 M_{*}^{16}}   \left[ H \left(2M_{*}^{2}\Gamma^{\prime}+2M_{*}^{2}\Lambda^{\prime} -4(M_{2}^{4})^{\prime} -9 H (\bar{M}_{1}^{3})^{\prime} \right)             \right]^{\prime}+\frac{M_{2}^{4}}{M_{*}^{8}} > 0 \, ,
\label{ReconstructedBound}
\end{align}
\end{widetext}
where a prime denotes a derivative with respect to $\ln a$. It is possible to express Eq.~\eqref{ReconstructedBound} in terms of the $\left\{\alpha_{M},\alpha_{B}, \alpha_{K} \right\}$ basis using Table II in Ref.~\cite{Kennedy:2017sof}. We shall now examine the implications of imposing Eq.~\eqref{ReconstructedBound} on the viable parameter space for different parameterizations of Horndeski gravity models.


\section{Constraining generic EFT parameterizations using positivity bounds}
\label{Sec.IV}
\begin{figure*}
\resizebox{0.325\textwidth}{!}{
\includegraphics{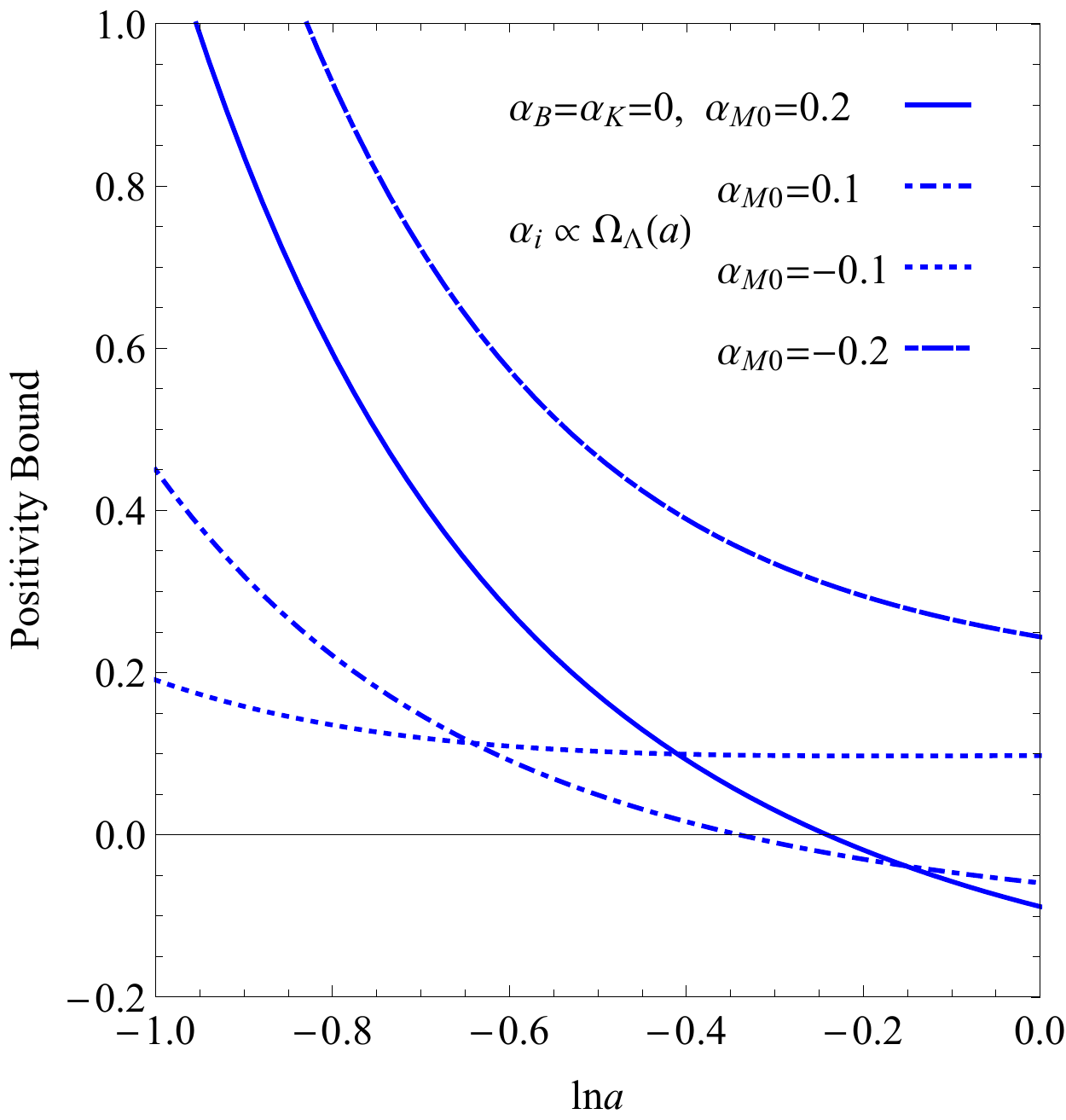}
}
\resizebox{0.325\textwidth}{!}{
\includegraphics{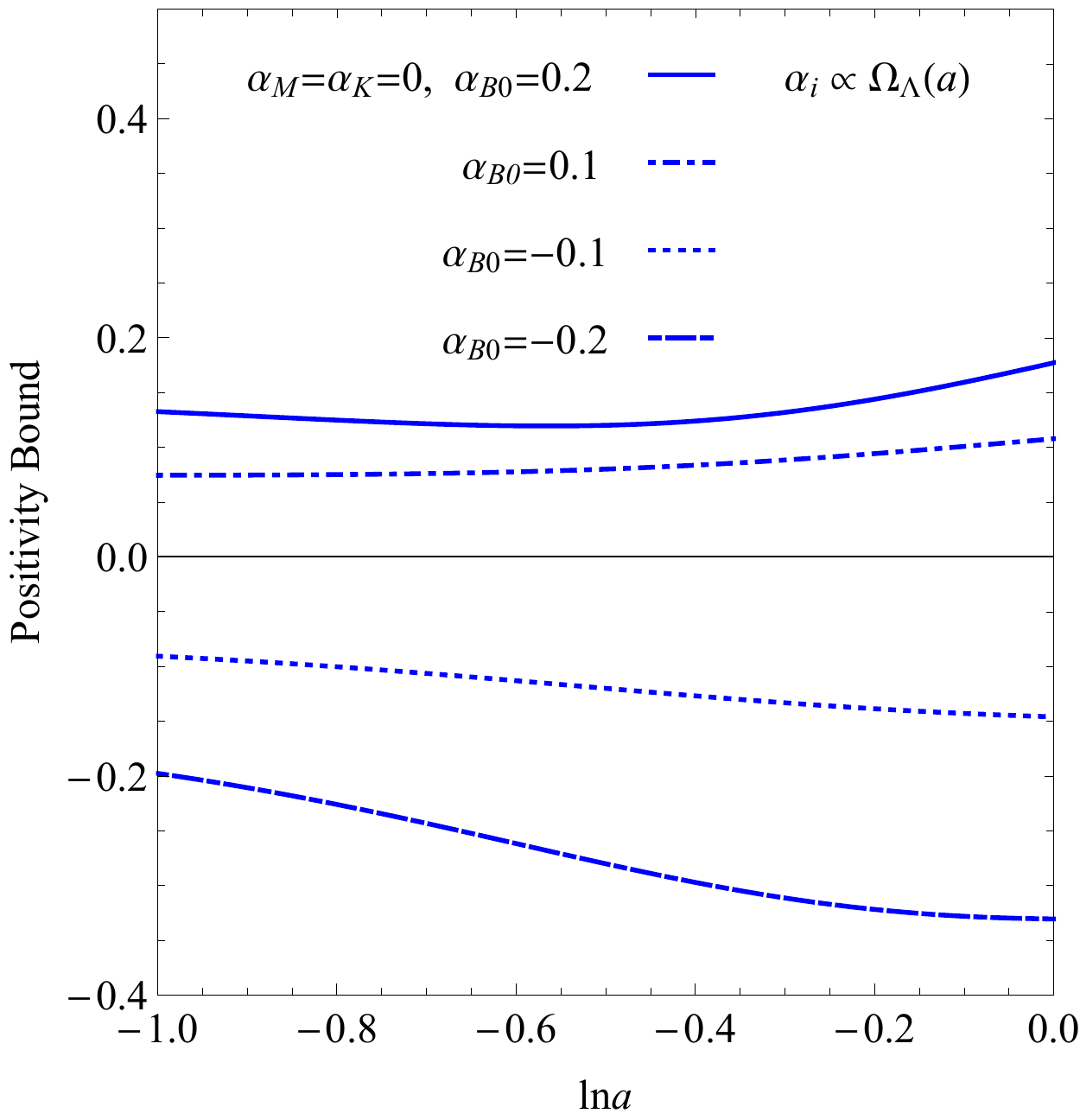}
}
\resizebox{0.333\textwidth}{!}{
\includegraphics{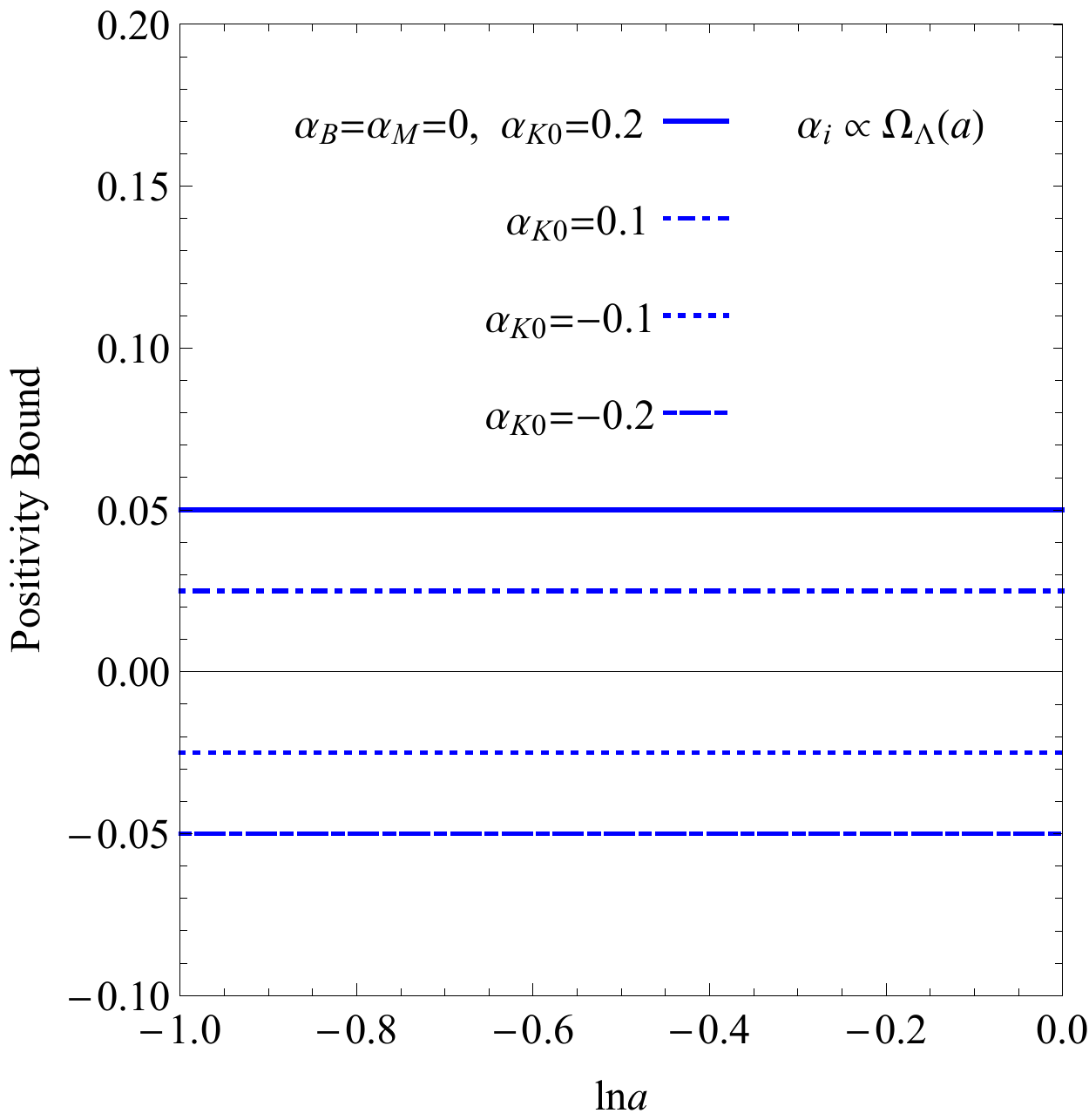}
}
\caption{
Time dependence of the left-hand side of the positivity bound in Eq.~\eqref{ReconstructedBound} for reconstructed Horndeski models with an underlying EFT parameterization of the form $\alpha_{i}= \alpha_{i0}\Omega_{\Lambda}(a)$ for sub-classes of models with only a non-zero $\alpha_{M}$ (left-hand panel), a non-zero $\alpha_{B}$ (middle panel) and a non-zero $\alpha_{K}$ (right-hand panel).
For a theory to be viable the curve must remain positive for the entire evolution in $\textnormal{ln}a$. One can see that for these particular sub-classes of models, the positivity bound is respected for $\alpha_{M}<0$, $\alpha_{B}>0$ and $\alpha_{K}>0$, with virtually no time dependence for models with only a non-zero $\alpha_{K}$. These plots are displayed in units with $M_{*}=H_{0}=1$. }
\label{Fig:OmegaDEtimeevolAmAbAk}
\end{figure*}
Having reviewed the physical arguments behind the positivity bound in Eq.~\eqref{eq:Pos1}, we shall now study the viable regions it implies for two commonly used phenomenological EFT parameterizations. 
In particular, we study theories reconstructed from parameterizations written in terms of the evolving fractional background dark energy density $\Omega_{\Lambda}(a)=H_{0}^{2}\Omega_{\Lambda 0}/H^{2}$ as
\begin{equation}
    \alpha_{i}(a)=\alpha_{i0}\Omega_{\Lambda}(a) \, ,
\label{param:1}
\end{equation}
or proportional to the scale factor $a(t)$ raised to some power $q$,
\begin{equation}
    \alpha_{i}(a)=\alpha_{i0}a^{q} \, ,
\label{param:2}
\end{equation}
with $i \in \left\{M,B,K\right\}$. Different choices are of course possible, one example being the inherently stable parameterization discussed in Sec.~\ref{Sec.V}.
The motivation for these particular forms stem from the sole requirement that the effects of any particular dark energy and modified gravity model giving rise to these EFT functions should become relevant during late-times to connect their evolution with late-time acceleration. 
By restricting to parameterizations of the form in Eqs.~\eqref{param:1} and \eqref{param:2} one is restricting to a subclass of underlying Horndeski theories. 
However, as these parameterizations are typically used in MCMC analyses constraining deviations from GR \cite{Ade:2015rim} 
it is of interest to determine what additional information can be provided by positivity bounds for this particular choice.
Another important question we can address is how sensitive the positivity bound is to the choice of parameterization and how it varies with the amplitudes of the EFT parameters $\alpha_{i0}$. 

\begin{figure*}
\resizebox{0.325\textwidth}{!}{
\includegraphics{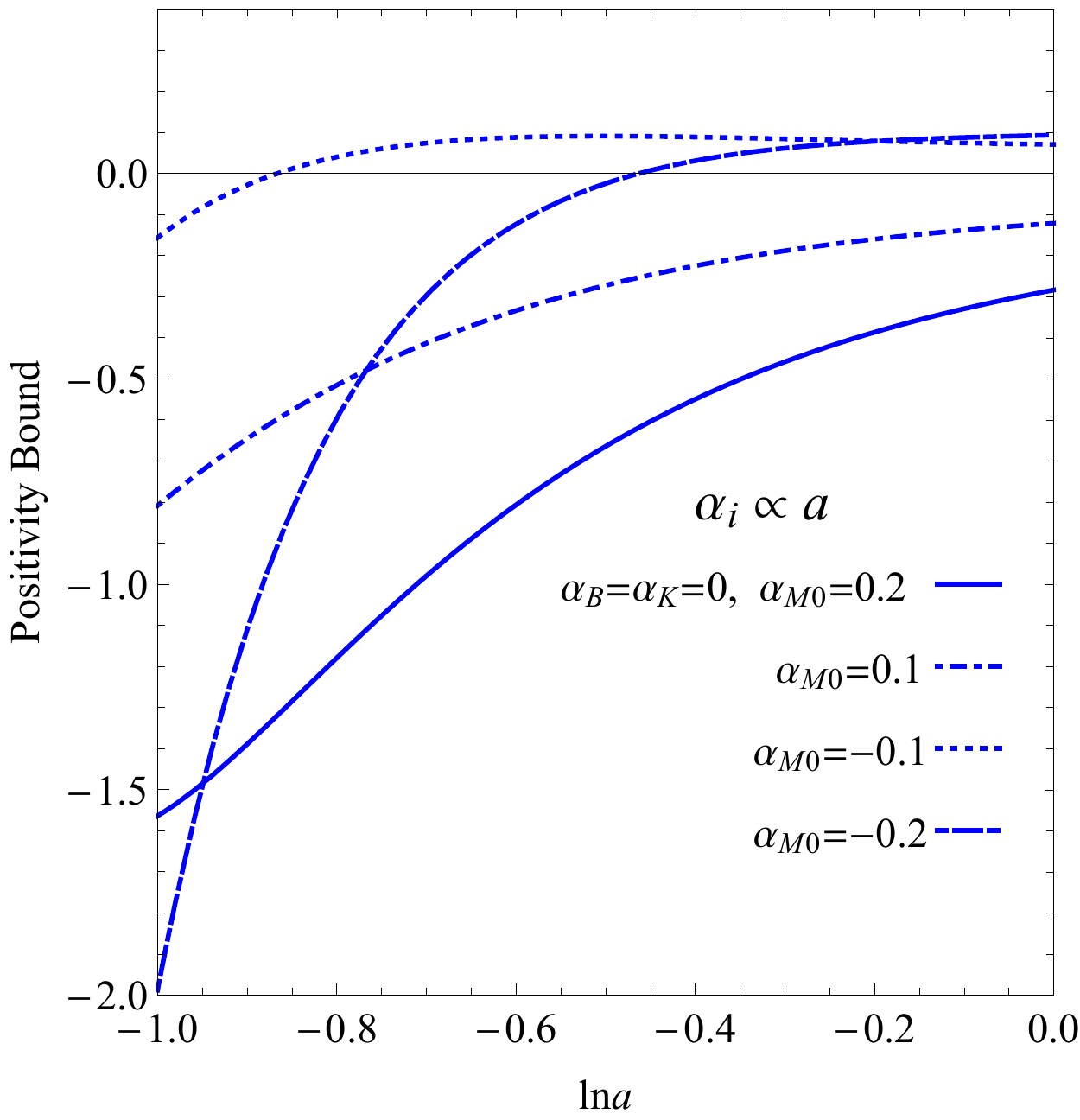}
}
\resizebox{0.325\textwidth}{!}{
\includegraphics{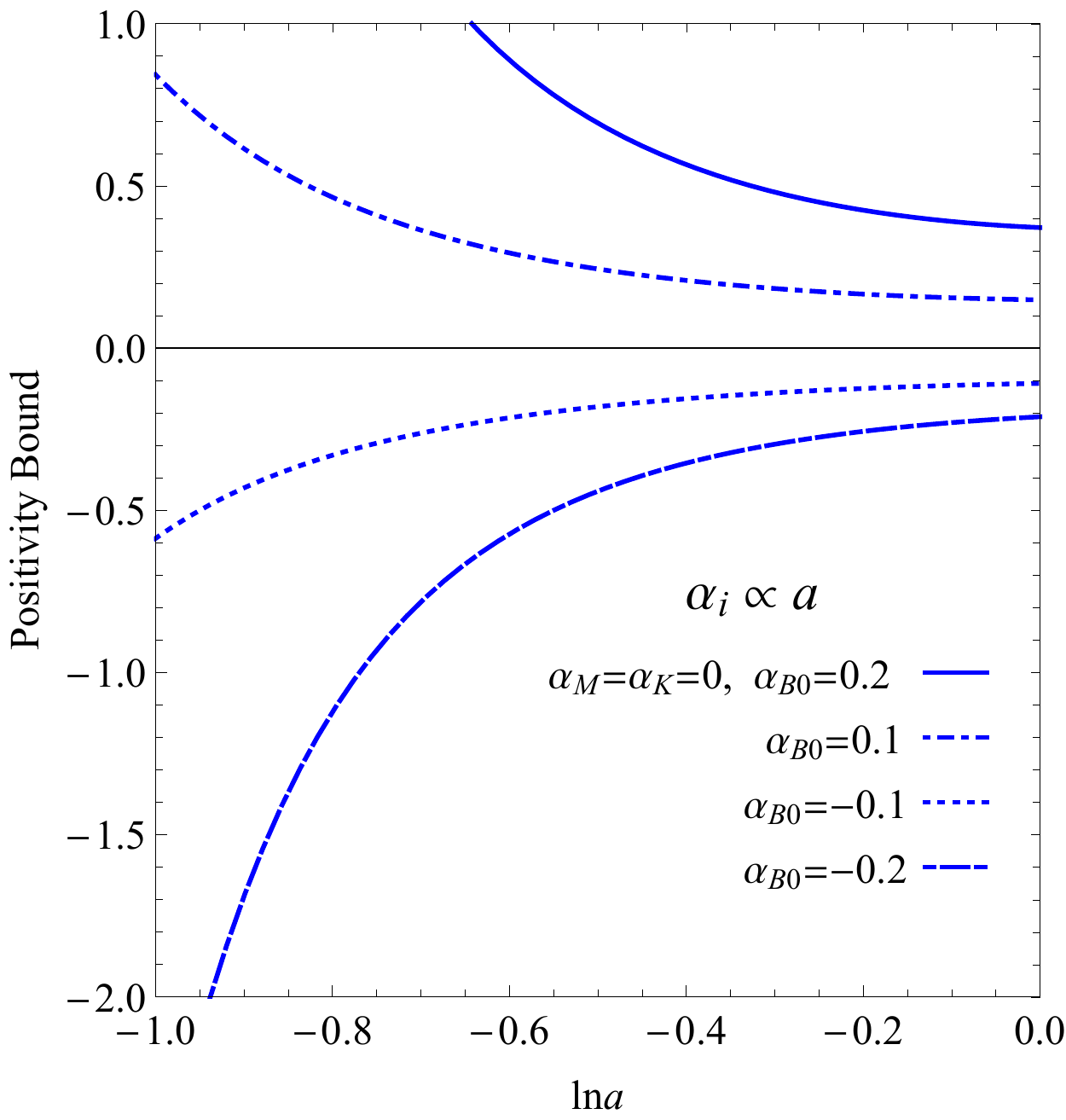}
}
\resizebox{0.325\textwidth}{!}{
\includegraphics{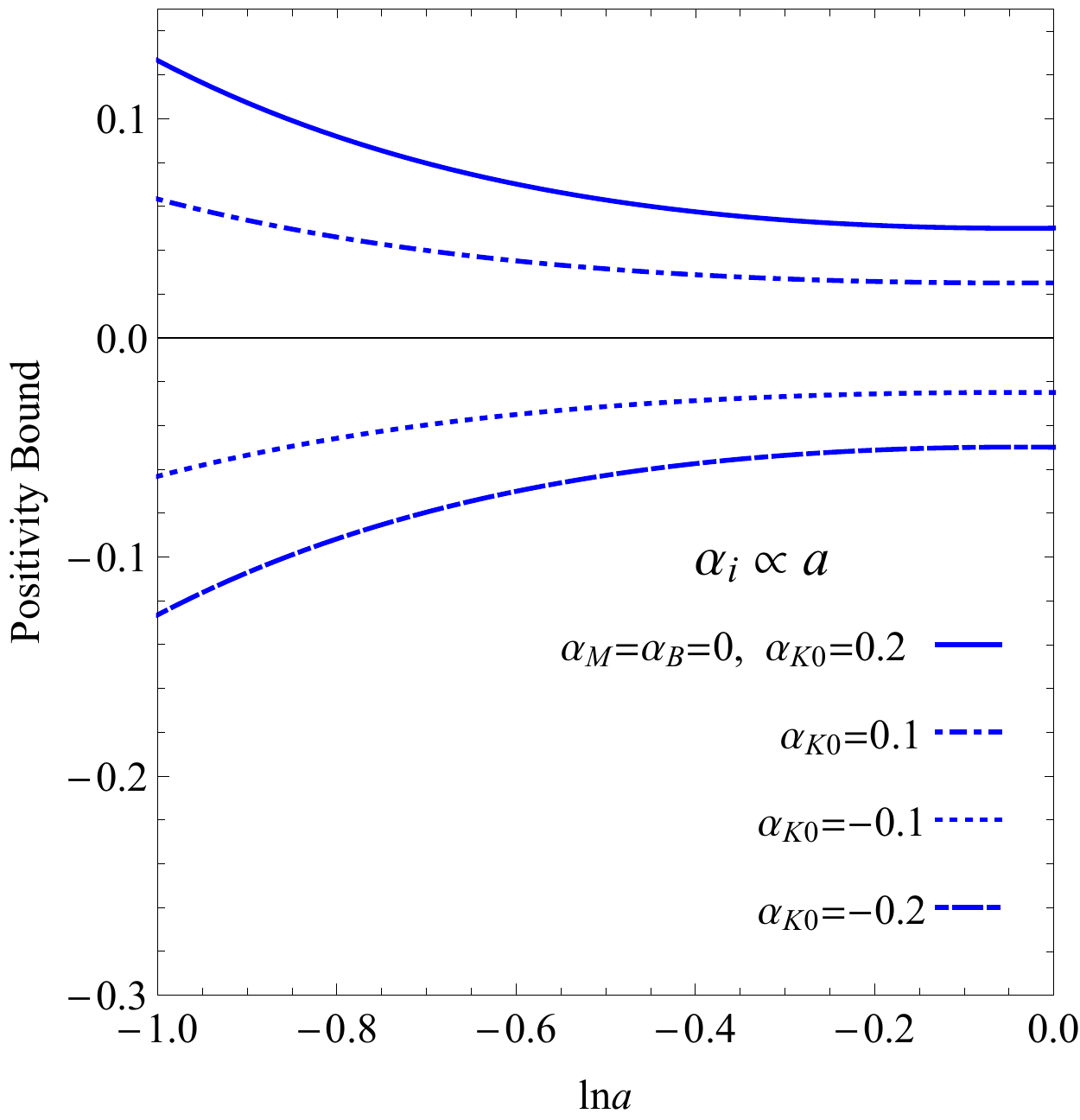}
}
\caption{
Evolution of the left-hand side of the positivity bound in Eq.~\eqref{ReconstructedBound}, in this case for parameterizations of the EFT functions in terms of the scale factor, Eq.~\eqref{param:2}. In this particular example we have chosen $q=1$. One can see in the left-hand panel that for the particular range of values chosen for $\alpha_{M0}$ there is no choice that will satisfy the positivity bound for all times. Restricting to later times, we see again the preference for $\alpha_{M}<0$ in order to satisfy the positivity bound. In the same manner, the middle panel indicates the positivity bound is satisfied in models with only a non-zero $\alpha_{B}$ as long as $\alpha_{B}>0$. 
The right-hand panel displays the evolution of the positivity bound for models reconstructed from a non-zero $\alpha_{K}$ only, again showing the preference for models with $\alpha_{K}>0$.
These plots are displayed in units with $M_{*}=H_{0}=1$.}
\label{Fig:ScaleFactortimeevolAmAbAk}
\end{figure*}

\subsection{Evolution proportional to dark energy density}

In Fig.~\ref{Fig:OmegaDEtimeevolAmAbAk} we present the evolution of the positivity bound for theories reconstructed from models with only a non-zero $\alpha_{M}$ (left-hand panel), a non-zero $\alpha_{B}$ (middle panel) and a non-zero $\alpha_{K}$ (right-hand panel) when parameterised in terms of the dark energy density, Eq.~\eqref{param:1}.
One can see from the left-hand panel with a varying $\alpha_{M}$ that the positivity bound favours models that have $\alpha_{M}<0$, as the chosen models with $\alpha_{M}>0$ eventually evolve to negative values at the present day.  
This also demonstrates a subtle point arising from the time dependence of the positivity bound when applying positivity priors in statistical sampling.
When constraining the EFT parameters with data-sets taken from a given redshift bin, it is important to check that every region of the parameter space within the bound allowed by positivity satisfies the positivity bound at all times until the present. 
Checking it at a single redshift is not enough. 
By using the reconstructed theory one can check that this holds for every point in the parameter space.
Similarly, the middle panel indicates that for the subset of models for which $\alpha_{M}=\alpha_{K}=0$ and with a non-zero $\alpha_{B}$ the positivity bound seems to favor those models with $\alpha_{B}>0$.
In addition, the right-hand panel displays the evolution of the positivity bound for models that only have a non-zero $\alpha_{K}$.
The positivity bound seems to favor $\alpha_{K}>0$ and has virtually no time dependence.
Of course these plots only give an indication of what we can learn by combining the reconstruction and the positivity bound in Eq.~\eqref{eq:Pos1}, being restricted to an arbitrarily chosen set of models.

In order to give a more comprehensive overview of the regions in the $\left\{\alpha_{M0},\alpha_{B0},\alpha_{K0}\right\}$ parameter space that satisfy the positivity bound we display in the top left-hand panel of Fig.~\ref{Fig:3d_sampled_plots} the three-dimensional viable region.
Each point in the parameter space spanned by $\left\{\alpha_{M0},\alpha_{B0},\alpha_{K0} \right\}$ is sampled between $\ln a=-1$ and $\ln a=0$ in intervals of $\Delta \ln a=0.05$ and the resulting plot displays those regions for which the positivity bound is satisfied for every sampled time.
One can see that the most straightforward way to ensure the positivity bound is satisfied is to have $\alpha_{M0}<0$ and $\alpha_{B0}>0$ which is in accordance with Fig.~\ref{Fig:OmegaDEtimeevolAmAbAk}. 
The apparent independence of the positivity bound on $\alpha_{K0}$ can also be seen, but this only holds for models with $\alpha_{M0}\lesssim 0$.
It is furthermore possible to have models which satisfy the positivity bound with $\alpha_{B0}\approx 0$, $\alpha_{M0}>0$ and $\alpha_{K0}>0$.

\subsection{Power-law evolution}

We shall now parameterize the EFT functions in terms of the scale factor as in Eq.~\eqref{param:2}, where we choose the power $q=1$, and examine the time dependence of the positivity bound for the corresponding reconstructed models.
The results are displayed in Fig.~\ref{Fig:ScaleFactortimeevolAmAbAk}.
As in the previous case, for models with only a non-zero $\alpha_{M0}$, the positivity bound favors models with $\alpha_{M0}<0$, but it cannot be satisfied for all time for the range of chosen $\alpha_{M0}$ values.
As for theories with only a non-zero $\alpha_{B}$, the positivity bound prefers models with $\alpha_{B0}>0$, and in models with only a non-zero $\alpha_{K}$, it prefers $\alpha_{K0}>0$, again with only minor time dependence.
Given this, one might determine that the conclusions drawn about the structure of the positivity priors are similar regardless of whether  Eq.~\eqref{param:1} or Eq.~\eqref{param:2} is chosen.
However, if we now examine the three-dimensional region plot of the values that satisfy the positivity bound, again sampled across times between $\ln a=-1$ and $\ln a=0$ at intervals of $\Delta \ln a=0.05$, we see in the top right-hand panel of Fig.~\ref{Fig:3d_sampled_plots} that the viable regions look different.
In contrast to the top left-hand panel there is now a stronger preference for models with $\alpha_{M0}>0$ and $\alpha_{B0}>0$.
This shows the priors placed on the EFT parameters from positivity bounds depend on the parameterization. 
Due to the non-trivial structure of the viable regions it is indeed difficult to make many generic statements about the regions where the bound is satisfied. 
This is perhaps surprising, as although different underlying theories are being tested with each parameterization, both have broadly the same phenomenological behaviour. 
Fortunately, in the next section we shall demonstrate how these problems can be avoided if one uses the inherently stable parameterization of the EFT parameters.

\section{Positivity bounds on theories reconstructed from an inherently stable EFT parameterization}
\label{Sec.V}
In the previous section we studied the regions of the EFT parameter space that respect the positivity bound for common phenomenological EFT parameterizations.
By choosing this set of parameters, even within the regions where the positivity bound is satisfied there is no guarantee that the theory does not possess gradient and ghost instabilities. 
To address this issue, we repeat here the same analysis as in Sec.~\ref{Sec.IV} using instead the inherently stable parameterization of Refs.~\cite{Kennedy:2018gtx, Lombriser:2018olq}, which parameterizes the model space in terms of the stability conditions themselves.
Viable regions satisfying the positivity bounds automatically satisfy the gradient and ghost stability conditions in this parameterization. 
Priors inferred on these inherently stable parameters therefore place a strong constraint by incorporating two independent theoretical requirements.
Concretely, the inherently stable basis is specified by four parameters $\left\{M^{2}, c_{s}^{2}, \alpha, \alpha_{B0}   \right\}$, the Planck mass squared, the scalar field soundspeed, the kinetic energy of the scalar field defined by $\alpha=\alpha_{K}+6\alpha_{B}^{2}$ and an initial condition for $\alpha_{B}$.
\begin{figure*}
\resizebox{0.35\textwidth}{!}{
\includegraphics{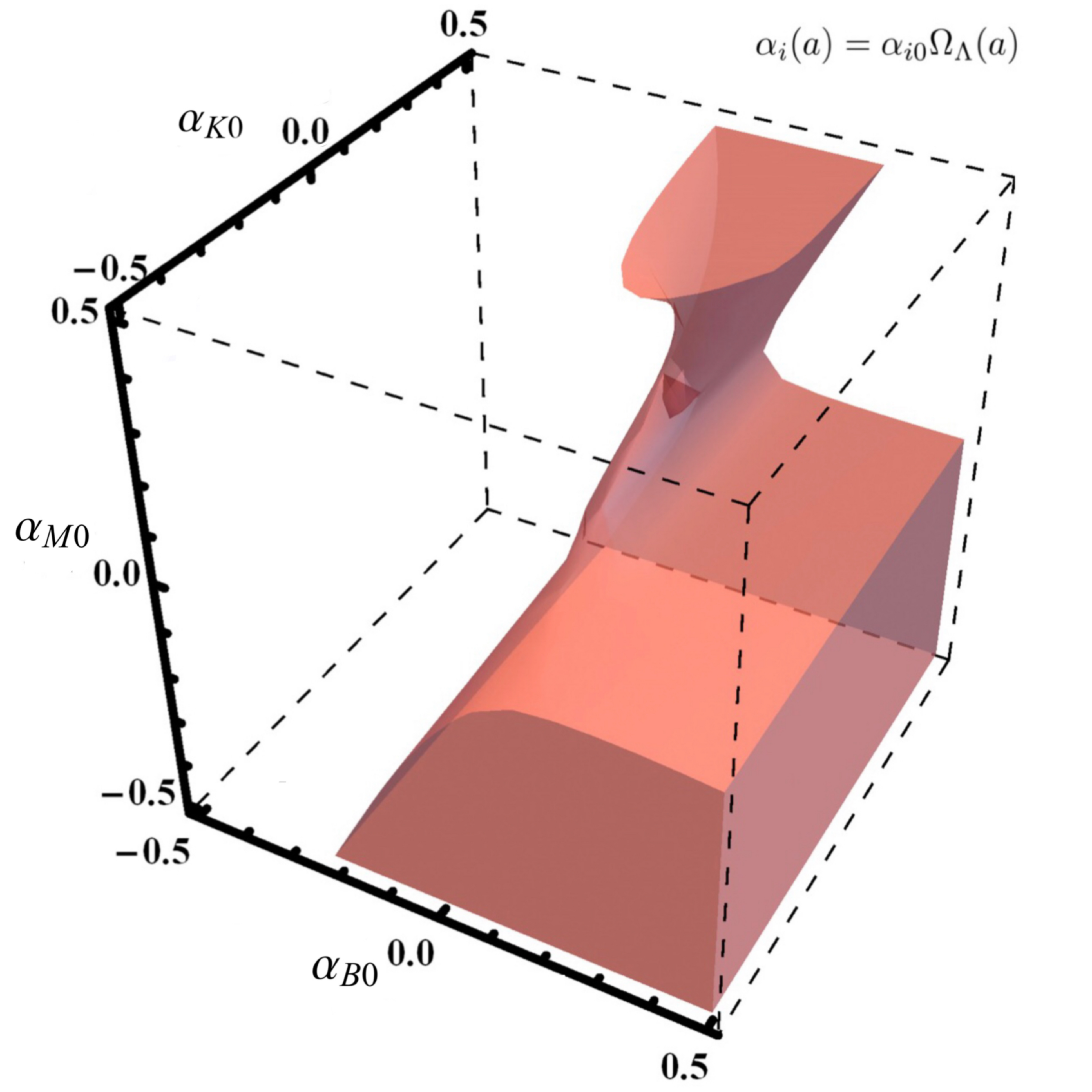}
}
\resizebox{0.37\textwidth}{!}{
\includegraphics{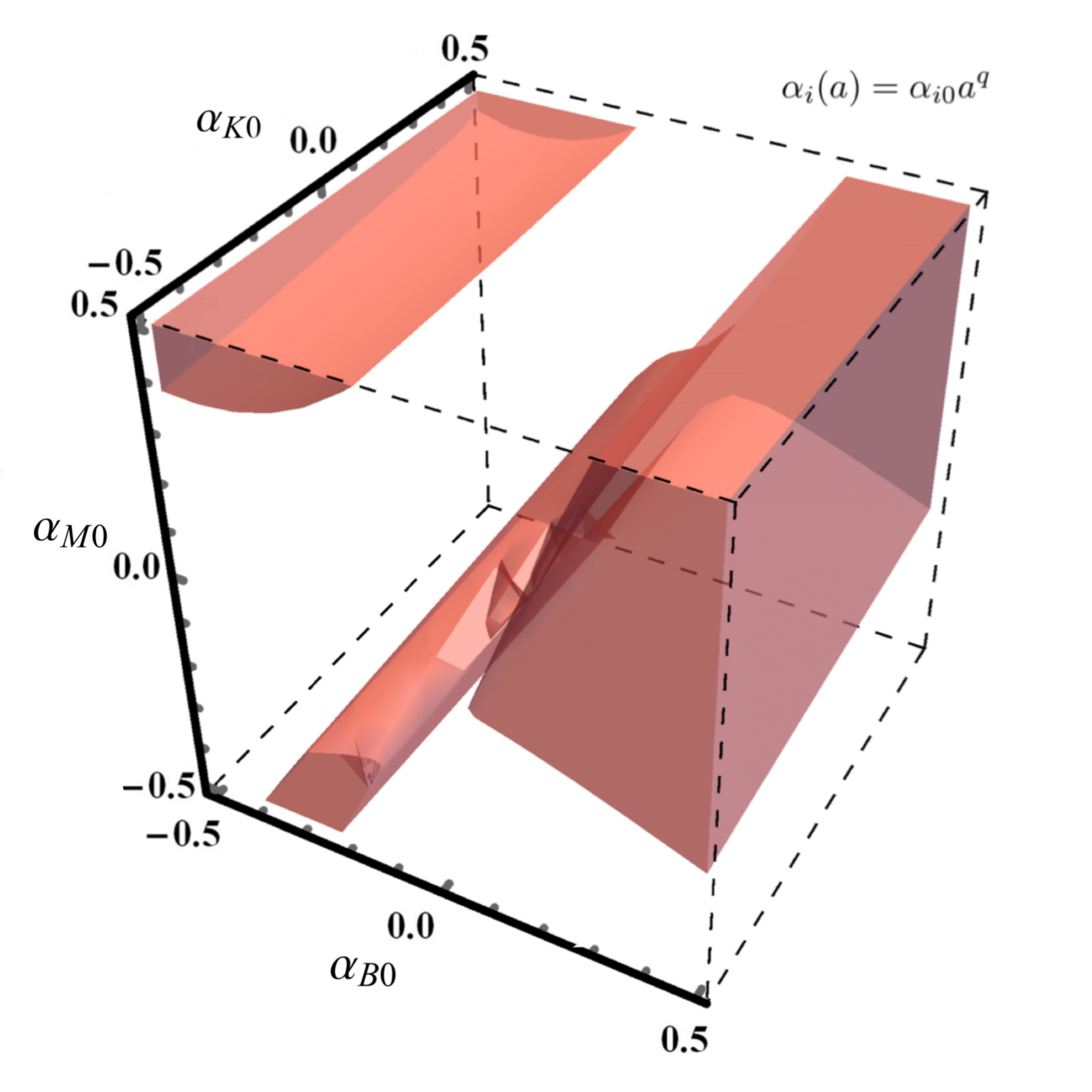}
}
\resizebox{0.32\textwidth}{!}{
\includegraphics{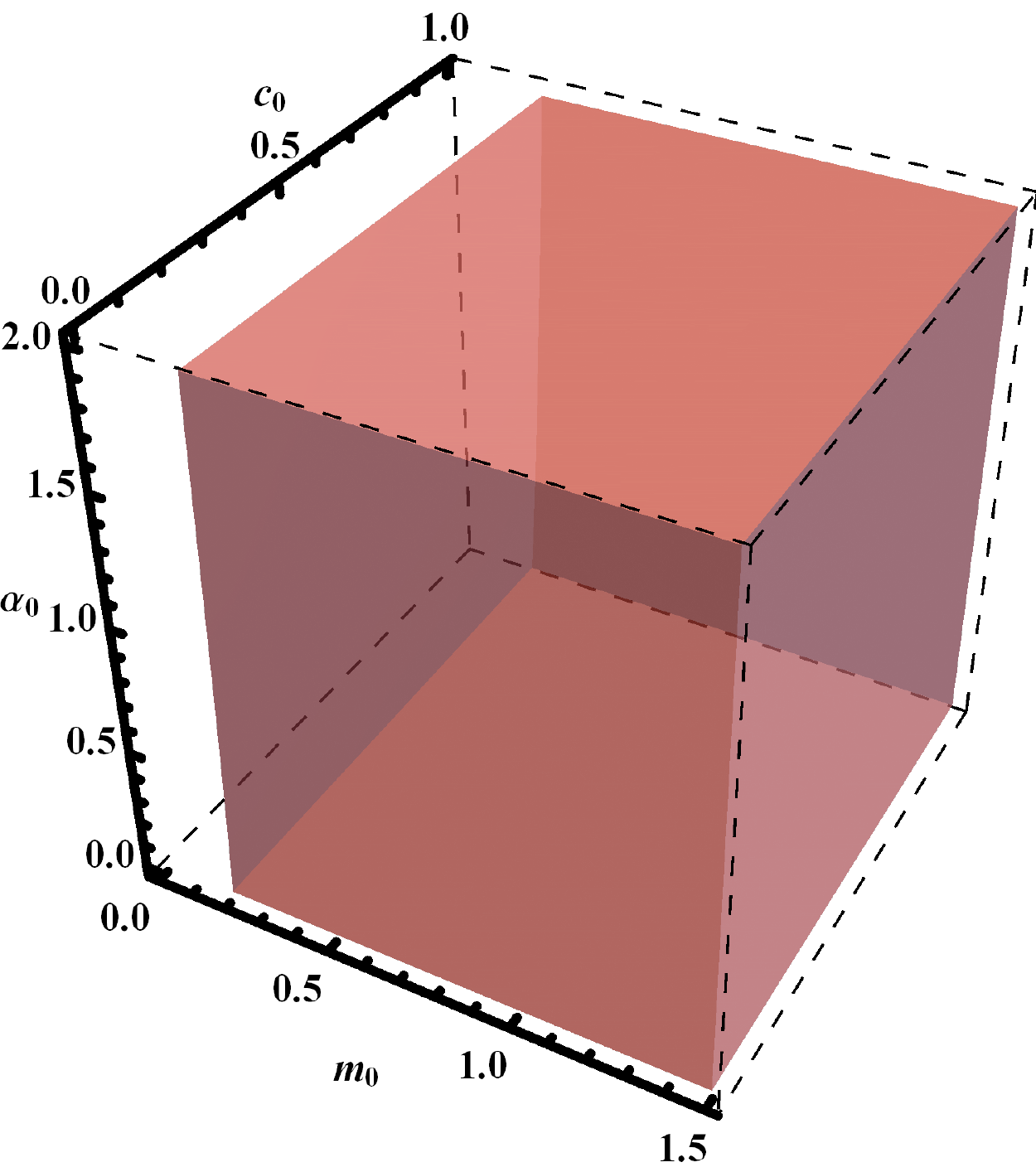}
}
\resizebox{0.32\textwidth}{!}{
\includegraphics{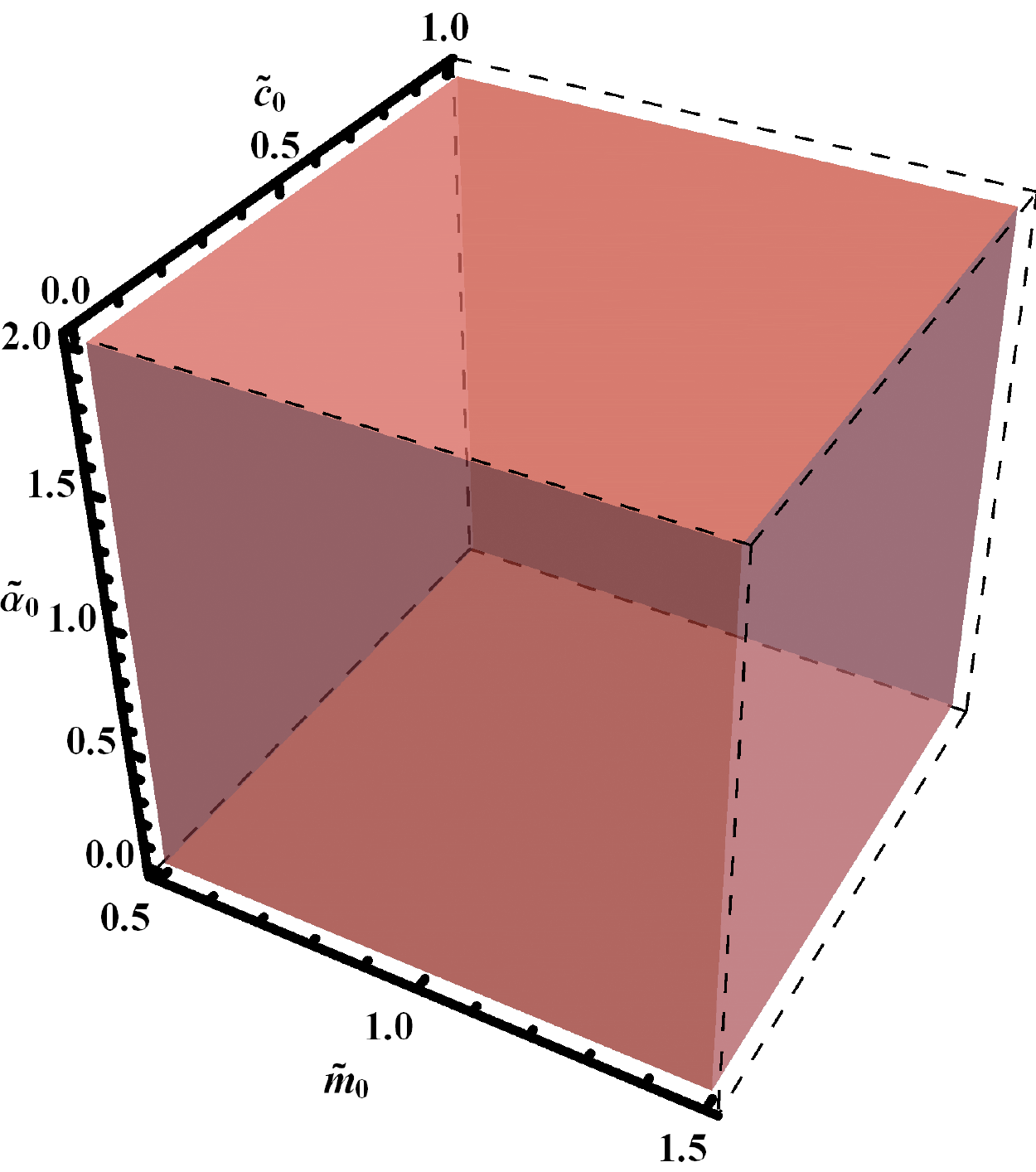}
}
\caption{The top row displays the three-dimensional regions in the $\left\{\alpha_{M0},\alpha_{B0},\alpha_{K0}\right\}$ parameter space which satisfy the positivity bound for models parameterized in terms of Eq.~\eqref{param:1} (top left-hand panel) and Eq.~\eqref{param:2} (top right-hand panel).
By examining cross sections of these surfaces a number of insights can be drawn. For example, inspecting the top left-hand panel one can see that for $\alpha_{B}=0$ the positivity bound is easier to satisfy for negative $\alpha_{M}$, unless $\alpha_{K}$ is positive. One can also see a general preference for a positive $\alpha_{B}$.
The qualitative difference between the top left-hand and the top right-hand panels clearly demonstrates the parameterization dependence of the positivity bound. Despite this there are a number of similarities. For example, in the absence of large deviations in $\alpha_{M}$ from zero the positivity bound prefers $\alpha_{B}>0$ and is rather insensitive to the value of $\alpha_{K}$ in both parameterizations.
By contrast, the bottom row displays the regions in the parameter space spanned by the inherently stable EFT functions whose reconstructed theory satisfies the positivity bound in Eq.~\eqref{eq:Pos1}. The lower left-hand panel illustrates the viable region when $M^{2}$, $c_{s}^{2}$ and $\alpha$ are parameterised in terms of the dark energy density. Only extremely small values of the present squared Planck mass $m_{0}<0.2$ are excluded, indicating it is difficult to violate the positivity bound when using this EFT basis. The situation is amplified when adopting a parameterization that satisfies the stability conditions for all time such as a Fermi distribution, where there is no region in the parameter space that violates the positivity bound, as can be seen in the lower right-hand panel.
Each plot is sampled at twenty points between $\ln a=-1$ and $\ln a=0$ and adopts units with $H_{0}=M_{*}=1$.}
\label{Fig:3d_sampled_plots}
\end{figure*}
As $\alpha_{B}$ and $c_{s}^{2}$ are related via a first-order differential equation
\begin{align}
c_{s}^{2}=&-\frac{2}{\alpha}\left[\alpha_{B}^{\prime}+(1+\alpha_{T})(1+\alpha_{B})^{2} \right. \nonumber \\
 &\left. -\left(1+\alpha_{M}-\frac{H^{\prime}}{H} \right)(1+\alpha_{B})+\frac{\rho_{m}}{2H^{2}M^{2}} \right] \, ,
 \label{soundspeed}
\end{align}
in order to find $\alpha_{B}$ given a choice of soundspeed, it is necessary to solve the resulting inverted second-order differential equation
\begin{equation}
 B'' - \left(1+\alpha_M-\frac{H'}{H} \right) B' + \left(\frac{\rho_m}{2H^2 M^2} + \frac{\alpha \: c_{s}^{2}}{2} \right) B = 0 \,,
\label{FindingaBfromsoundspeed}
\end{equation}
with 
\begin{equation}
    1+\alpha_{B}\equiv\frac{B^{\prime}}{B} \, .
\end{equation}
In order to satisfy the no-ghost and no-gradient instability conditions we require
\begin{equation}
    M^{2}>0 \, , \, 0<c_{s}^{2}<1 \, , \, \alpha>0 \, .
\end{equation}
If we then choose parameterizations that respect these bounds the constraints arising from observational data will automatically satisfy the stability conditions.

Let us start by again parameterising each of these EFT parameters in terms of the dark energy density
\begin{align}
    M^{2}&=1+\left(m_{0}-1\right)\Omega_{\Lambda}(a)/\Omega_{\Lambda0}\, , \label{ISPOmega1} \\
    c_{s}^{2}&=1+\left(c_{0}-1\right)\Omega_{\Lambda}(a)/\Omega_{\Lambda0}\, , \label{ISPOmega2} \\
    \alpha&=1+\left(\alpha_{0}-1\right)\Omega_{\Lambda}(a)/\Omega_{\Lambda0}\, ,  \label{ISPOmega3}
\end{align}
where $m_{0}$, $c_{0}$ and $\alpha_{0}$ are defined such that they are values of the Planck mass, soundspeed and kinetic energy today. 
Using the reconstruction we can derive a set of $G_{i}(\phi,X)$ functions that give rise to this set of EFT functions, and subsequently then use these in the positivity bound to examine the overlap between theories that satisfy the stability conditions and those that satisfy the positivity bound. 
Displayed in the lower left-hand panel of Fig.~\ref{Fig:3d_sampled_plots} is the region in the $\left\{m_{0},c_{0},\alpha_{0}   \right\}$ parameter space that satisfies the positivity bound. 
It is clear that the positivity bound is only violated for rather drastic deviations in the squared Planck mass.
It also shows no dependence on the soundspeed and the kinetic energy.
Note this is in sharp contrast to the parameterizations examined in the previous section, where small deviations in the EFT parameters could easily move to a region that violated the positivity bound.

It is again natural to address whether this result was dependent on the particular parameterization that we adopted for the inherently stable basis.
This is perhaps a greater issue with the inherently stable parameterization as it is clear that, while the parameterizations in Eqs.~\eqref{ISPOmega1} to \eqref{ISPOmega3} respect the stability conditions at all times up until the present day, one may worry that $c_{s}^{2}$ evolves to values greater than one in the future.
We shall therefore perform the same analysis but with the following more physically realistic choice for the time evolution of the EFT functions
\begin{align}
    M^{2}&=1+\frac{\Tilde{m}_{0}-1}{1+e^{-\textnormal{ln}a}}\, , \label{ISPFermi1} \\
    c_{s}^{2}&=1+\frac{\Tilde{c}_{0}-1}{1+e^{-\textnormal{ln}a}}\, , \label{ISPFermi2} \\
    \alpha&=1+\frac{\Tilde{\alpha}_{0}-1}{1+e^{-\textnormal{ln}a}}\, ,  \label{ISPFermi3}
\end{align}
where we have kept the time dependence explicitly in terms of $\ln a$ in order to keep the equivalence with the Fermi distribution transparent.
With this choice, the inherently stable parameters are guaranteed to satisfy the stability bounds for all time as long as $\Tilde{m}_{0}\geq 0$, $\Tilde{\alpha}_{0} \geq 0$ and $0\leq\Tilde{c}_{0}\leq1$, smoothly transitioning from a constant value set to one at early times to a constant value in the future specified by $m_{0}$, $\alpha_{0}$ and $c_{0}$.
As we can see in the lower right-hand panel of Fig.~\ref{Fig:3d_sampled_plots} every sampled point satisfies the positivity bound, where again the sampling has been performed between $\ln a=-1$ and $\ln a=0$ at intervals of $\Delta \ln a=0.05$. 
It is perhaps not surprising that the likelihood of satisfying the positivity bound increases even further when using a physically more realistic parameter choice that satisfies the no-ghost and no-gradient instability conditions for all time.
Compared with the results of Sec.~\ref{Sec.IV} one can see that by using a parameterization that avoids ghost and gradient instabilities, the positivity bound is more likely to be satisfied.
Intuitively this makes sense, as one would expect a theory that is theoretically healthy with respect to one condition to be more likely to be theoretically healthy with respect to another condition, although this was by no means guaranteed.

We conclude that this provides extra motivation for making use of the inherently stable parameterization for the EFT functions in upcoming surveys, as they provide the additional benefit of naturally satisfying positivity bounds.

\section{Conclusions}
\label{Sec:conclusions}
We are about to enter an era in which an unprecedented amount of cosmological data will become available from the next generation of surveys such as LSST~\cite{Ivezic:2008fe} and Euclid~\cite{Laureijs:2011gra}.
These data-sets will provide a testing ground for GR in the cosmological regime, with the additional prospect of shedding some light on the nature of cosmic acceleration. 
To optimally exploit the data for this purpose, it is important to study the potential impact theoretical arguments can have in placing priors on the parameter space that will be sampled.  

In this work we have combined the reconstruction approach of Refs.~\cite{Kennedy:2017sof, Kennedy:2018gtx} and the positivity bound derived for Horndeski theory in Ref.~\cite{Melville:2019wyy} to study the impact that respecting this bound has on the EFT parameter space.
If a theory fails to satisfy the positivity bound, it indicates that the theory at higher energies fails to adhere to important properties of quantum field theory such as causality, locality and crossing symmetry, even if the theory itself respects all of these conditions at low energies.
Their utility lies in the fact that knowing the precise form of the UV-completion is irrelevant to apply the arguments.
Treating Horndeski theory itself as some low-energy effective theory of an as yet unknown UV-completion, Ref.~\cite{Melville:2019wyy} derived the conditions that the $G_{i}$ functions must satisfy in order to respect the positivity bound.
The reconstruction method enables the translation between bounds derived for the $G_{i}$ functions to bounds on the EFT parameters themselves for different choices of phenomenological parameterizations.

In Sec.~\ref{Sec.IV} we studied the resulting constraints on the EFT parameters for two parameterizations commonly adopted in the literature for testing dark energy and modified gravity models.
These parametrizations tie their evolution to the dark energy density or the scale factor raised to some power.
We have found that the resulting reconstructed theories only satisfy the positivity bounds in non-trivial regions of the parameter space. 
Minor deviations in the amplitude of these EFT functions from their GR value can reconstruct a theory which fails to satisfy the positivity bound, and so one has to be careful in placing positivity priors when adopting this choice of parametrization.
Applying positivity priors when using these parameterizations may lead to tight constraints on the EFT functions. However, these constraints are simply an artificial artefact of a poor choice of EFT basis.
By contrast, in Sec.~\ref{Sec.V} we found that if the inherently stable basis of dark energy and modified gravity models is adopted instead, it is much more difficult for the resulting reconstructed models to violate the positivity bound.
With the parameterizations we adopted for the inherently stable EFT basis, we found it was either only possible to violate the positivity bounds in physically unrealistic regions of the parameter space, or not at all.
This gives extra motivation to make use of the inherently stable basis for upcoming surveys as by doing so, one is naturally avoiding gradient and ghost instabilities as well as sampling theories that satisfy the positivity bound.

A simple extension of this work would be to study even more parameterizations of the EFT functions to sample different regions of the theory space. 
A further extension will be to analyse how these bounds differ when including information from nonlinear scales, with a differing evolution of the kinetic term of the scalar field $X$ from its background value. %
It may then be possible to start constraining the nonlinear EFT functions from positivity bounds using the extended reconstruction method of Ref.~\cite{Kennedy:2019nie} 

%
\acknowledgments
%
We would like to thank Johannes Noller and Francesco Riva for useful discussions.
We also thank the former for comments on a preliminary draft of this work.
J.K.~and LL.~acknowledge the support of a Swiss National Science Foundation Professorship grant (No.~170547).
Please contact the authors for access to research materials.
%

\bibliographystyle{JHEP}
\bibliography{library}

\end{document}